\documentclass[aps,prb,twocolumn,showpacs,floatfix,pdf]{revtex4}

\usepackage{amsmath}    
\usepackage{graphicx}   
\usepackage{verbatim}   
\usepackage{color}      
\usepackage{subfigure}  
\usepackage{hyperref}   
\usepackage{epstopdf}
\usepackage{ulem}

\begin{document}
\title{Green's functions and equations of motion in correlated many body systems}

\author{Qingguo Feng}
\email[]{qingguo.feng@ifm.liu.se}
\affiliation{Department of Physics, Chemistry and Biology (IFM), Link\"{o}pings University, S-58183 Link\"{o}ping, Sweden.}


\begin{abstract}
In this paper some new physical notations are given for the Green's functions and equations of motion (EOM)
in many body physics with the concept of quasiparticles. It shows how the many body correlations existing
in many body systems can be taken into account through the derivation of EOMs self-consistently. The scheme is compared
with the expansion of Feynman diagrams. Finally some further discussions are given for the unique properties of EOM techniques in helping to model a system and the interactions self-consistently.
\end{abstract}


\maketitle

\section{Introduction\label{sect1}}
The concepts and techniques of Green's functions have been developed and widely used in the studies of many body physics.\cite{mahan} Once the Green's function is known, most of the information of the quantum system can be obtained. Along with the Green's function, the equation of motion (EOM) techniques have also been developed. For example, different EOM methods have been constructed to solve the impurity problems and to solve the many body system directly \cite{Lacroix1,Lacroix2,Czycholl85,GrosEOM,Luo99,JK05,ZhuEOM1,FZJ09,FO11a,FO11b,FO11c,poland12} or with the dynamical mean field theory.\cite{DMFTRMP96,DMFTRMP06}
In its application, the EOM method is one of the basic, powerful method to find the Green's functions. Moreover, it is also a mono-decisive method to find the solution. Under various physical approximations it gives a set of closed integral equations which can be solved numerically. Therefore, the EOM method is usually applied and assumed to be an approximate numerical method. However, Comparing with quantum Monte Carlo (QMC) method,\cite{HF-QMC} the exact diagonalization (ED) method,\cite{ED,ED2}
and the continuous time quantum Monte Carlo (CT-QMC) method,\cite{CTQMC,CTQMC2} the EOM method has its advantage that it is not only a numerical method but also a part of many body theory that can help to develop and examine the validity of an existing or a newly emerging theory.

In this paper we will give a few new physical elaboration of the Green's functions and EOMs, and then discuss the relation between EOMs and Feynman diagrams. Finally we discuss with example to show how the EOMs can verify a Hamiltonian so as to model the system in a self-consistent way.

\begin{figure}[tb!]
\includegraphics[angle=0,width=0.99\linewidth]{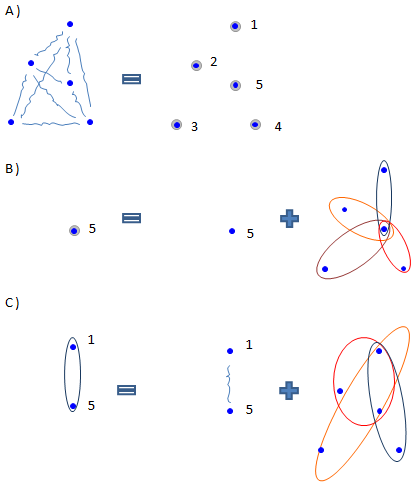}
\caption{(Color online) Illustration of an arbitrary correlated many body system. A) A many body system with interactions can be considered as a system of quasiparticles. The wave-like lines label the interactions between any two particles.  B) A quasiparticle can be assumed as sum of a 'naked' particle and the relative quasi-two-body interactions, where each ellipse circled area labels a quasi-two-body interaction.  C) A quasi-two-body interaction is similarly a sum of a 'naked' two-body interaction and the corresponding quasi-three-body interactions, where each ellipse circled area labels a quasi-three-body interaction. Here in B and C only one quasi-interactions is illustrated. Other same order quasi-interactions and those with higher order (e.g., quasi-Four-body interactions) can be similarly plotted.
\label{fig1}}
\end{figure}

Now let us imagine an arbitrary correlated many body system, as shown in Fig.~\ref{fig1}. The particles can be any correlated objects. But here in our interest the particles will be electrons. All these electrons can be either on one impurity site which will then be a single impurity problem, or on different sites as a lattice problem.
In a general situation (or presumption) each particle will interact with all the other members. If for some cases there is no interaction between any pair of members, one can still assume that there is interaction between these two members but the interaction strength for this 'nonexist' interaction is zero. Thus all the interactions in the whole system can be written as a Hamiltonian ${\cal H}$. From the many body theory, the complete information of each particle can be described with a single particle Green's function.
However, one should note that here the particles are not free particles but with interactions, i.e., they are quasiparticles. Therefore, the introduced single particle Green's function should be a sum of one term $G^{1,naked}_f$ corresponding to the 'naked' particle plus the contributions from the quasi-two-body interactions, as illustrated in (b) of Fig.~\ref{fig1}. We assume that all the three-body interactions, four-body interactions and other higher order interactions have been included in this quasi-two-body Green's function. Continue this procedure, this quasi-two-body Green's function equals a 'naked' two-body interaction plus a quasi-three-body Green's function as shown in (c) of Fig.~\ref{fig1}. One should note that there is no isolated two-body interaction or any isolated $n$-body interaction because one $n$-body-interaction associated members will always interact simultaneous with other members outside these $n$ members, i.e., the effect of $n$+$1$ body interactions and even higher order interactions will be expressed in this $n$-body interaction. Repeat this procedure again and again, one can clearly observe that all the higher interactions/correlations will be covered in the quasi-interaction with one order lower, and then finally covered by the quasi-single-particle, as presented in the following equations,
\begin{eqnarray}
G^{1,quasi}_{j}&=&G^{1,naked}_{j}+\sum_{j',j'\neq j}G^{2,quasi}_{jj'},\label{eq1}\\
G^{2,quasi}_{jj'}&=&G^{2,naked}_{jj'}+\sum_{\substack{
   j'' \\
   j''\neq j ~\&~ j''\neq j'
  }}G^{3,quasi}_{jj'j''},\label{eq2}\\
&\vdots&\nonumber
\end{eqnarray}
where the indices $j,j',j''$ label difference particles. From the above considerations, the achieved single-particle Green's functions should be theoretically exact.

To connect to our theory of EOMs, the above formulae just correspond to the derivation of higher and higher order EOMs as shown in Ref.~\onlinecite{FZJ09,FO11a,FO11b,FO11c},
\begin{eqnarray}
\omega \ll f_{\sigma};f^{\dag}_{\sigma}\gg&=&\langle [f_{\sigma},f^{\dag}_{\sigma}]_+\rangle+\ll [f_{\sigma},{\cal H}];f^{\dag}_{\sigma}\gg,\label{eq3}\\
\omega \ll\hat{n}_{f\sigma'}f_{\sigma};f^{\dag}_{\sigma}\gg&=&\langle [\hat{n}_{f\sigma'}f_{\sigma},f^{\dag}_{\sigma}]_+\rangle\nonumber\\
&&+\ll [\hat{n}_{f\sigma'}f_{\sigma},{\cal H}];f^{\dag}_{\sigma}\gg,\label{eq4}\\
&\vdots&\nonumber
\end{eqnarray}
where we have used the double time temperature-dependent retarded Green's functions in Zubarev's notation,\cite{Zubarev} $f^{\dag}_{\sigma}$ and $f_{\sigma}$ are the corresponding creation and annihilation operators of fermions with $\sigma$ and $\sigma'$ labeling different spin channels.
Comparing Eqs.~\eqref{eq1},\eqref{eq2} with \eqref{eq3},\eqref{eq4}, one can draw a conclusion that the procedure of deriving higher and higher order EOMs can distinguish the contributions of the interactions among different number of particles. Therefore, in EOM method it is easy to extract the higher order many-body correlation effects which may be very important in explaining the unconventional behaviors of quantum systems. 
For example, if the superconductivity does relate to the pair formed with one electron and one hole left by another electrons, the higher order long range correlations such as $\hat{n}_{i\sigma}f^{\dag}_{j\sigma'}f_{i\sigma'}$ must have made great contributions.

The above EOMs are theoretically exact because each Green's function has taken into account all the higher order effects. However, when the number of particles in this correlated system increases, the system will become more and more complicated due to the increasing number of EOMs and the number of Green's functions, which will make it difficult to explore the system. In some situations the EOMs are resolvable, and in some cases they are not. For the latter cases, one simple and convenient way is to include the most important interactions only. Since a higher order interaction will be somehow a minor correction to the interaction with one order lower (here one should note that the higher order one can {\it never} be stronger than the interaction with one order lower because the interaction with one order lower will cover this higher order one), which makes it possible to truncate at one order according to the requested accuracy of one calculation. If one truncate this procedure at $n$ order, a $n$+$1$ order quasi-particle Green's function will be decoupled, which means that only the $n$-body interactions have been taken into account exactly. All the higher order interactions beyond $n$ order will be treated with mean field approximation which will cause the lose of some features of those correlations with order higher than $n$. The larger the truncation order $n$ is, the more accurate the quasi-single-particle Green's function is. At this point the EOM method is indeed an approximation method. However, one can always get satisfactory precision by increasing the truncation order though it is a hard work to derive the EOMs and made the decoupling. Anyway, physical approximations are always used in the studies of many body systems, e.g., mean field approximation between orbitals, the tight binding approximation between sites, or even the approximation with the next or next next nearest neighbors. If one increase the truncation order of EOMs higher than these physical approximations, the numerical error introduced by the decoupling can be definitely negligible.

\begin{figure}[tbh]
\includegraphics[angle=0,width=0.95\linewidth]{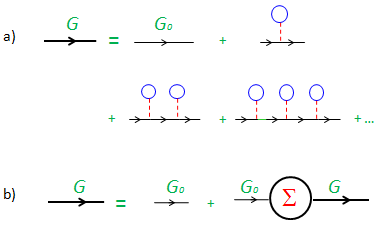}
\caption{(Color online) Illustration of diagram expansion. $G$ labels a Green's function of a quasiparticle and $G_0$ labels Green's function of bare particle. a) The expansion of Feynman diagrams. Here only Hartree-type terms are illustrated and complete diagrams should have Fock-type terms. b) Diagrammatic description of Dyson equation, where $\Sigma$ means self-energy. \label{fig2}}
\end{figure}

Next we will compare the EOMs to the diagram expansion. The usual expansion of the Feynman diagrams is as shown in Fig.~\ref{fig2},\cite{qft} where we only illustrated the Hartree-type diagrams.
From the expansion, one can observe that it is very similar to the EOM method. The higher order Feynman diagrams actually just correspond to the contributions of those higher order correlation effects shown in EOMs. However, in physical meaning these two methods are quite different. In EOM method the derivation of EOMs is actually equivalent to solve the Schr\"{o}dinger equation (or Dirac equation if relativistically) and in each derivation the whole Hamiltonian acts, while diagram expansion is from perturbation. In EOM method each higher order correlation effects will influence and contribute to the quasi-particle-correlations of one order lower, while in diagram expansion it is a simple summation of all the orders of diagrams. Therefore, it will lead to another difference that each higher Feynman diagram can be calculated independent to other diagram, while all the Green's functions in EOMs are unknown and can only be known by solving the equations self-consistently. Moreover, in achieving the Dyson equation,
\begin{eqnarray}
G=G_0+G_0\Sigma G,
\end{eqnarray}
obviously the higher order terms have also been decoupled in rewriting all the higher order terms as $G_0\Sigma G$.

From the above discussions, one can learned that, different to QMC and ED methods, the EOM technique is not only a numerical method to solve the eigenvalue problem. It can help to explore the many body system, or to examine a many body theory. Given a Hamiltonian, if the system is well modeled, {\it in the derivation of EOMs all the possible quantum many-body correlations will be naturally reproduced}. Otherwise there must be some problems or the theory is not self-consistent. For EOM method, the Hamiltonian will only consist of the most basic interactions (or the actual interactions). Those higher order correlations whose origin are owing to these basic interactions will appear in the higher order EOMs but not show in Hamiltonian. While for a pure numerical method solving eigenvalue problem usually the considered correlations have to be written in the Hamiltonian, which is only one alternative way to study the correlation effects because there they can not be self-consistently involved. If one want to study more higher correlations, more terms will be added into Hamiltonian. In this sense these "numerically exact" methods can realize and find the solution for a give Hamiltonian, and are only numerically exact for the given Hamiltonian. Definitely they will not exactly reflects the real system if an ill Hamiltonian has been given.

For example, for any a system, the most general model should cover all the possible cases and take into account all the possible physical effects. Therefore, for a multiorbital system the intersite interorbital hoppings and on-site interorbital hoppings will be introduced as done in Ref.~\onlinecite{FO11a}-\onlinecite{FO11c},
\begin{eqnarray}
{\cal H}&=&-\sum_{ijlm\sigma,i\neq j}t_{ijlm}f^{\dag}_{il\sigma}f_{jm\sigma}
+\sum_{il}U_{ll}\hat{n}_{il\uparrow}\hat{n}_{il\downarrow}\nonumber\\
&&+\sum_{ilm\sigma\sigma',l<m}U_{lm\sigma\sigma'}\hat{n}_{il\sigma}\hat{n}_{im\sigma'}\nonumber\\
&&+\sum_{ilm\sigma,l<m}\big(V'^{\ast}_{lm\sigma}f^{\dag}_{im\sigma}f_{il\sigma}^{~}+V'_{lm\sigma}f^{\dag}_{il\sigma}f^{~}_{im\sigma}\big).\label{eq6}
\end{eqnarray}
The first term on the right hand side (RHS) is the intersite hoppings. The second and third terms are the onsite intraorbital and interorbital Coulomb interaction terms. The last two terms are the onsite interorbtial single electron hoppings. The above Hamiltonian is introduced based on the following presumption. When there is no external field and potential, the system should only have the kinetic energy and Coulomb interactions, i.e.,
\begin{eqnarray}
{\cal H}&=&E_k+E_{Coulomb}.\nonumber
\end{eqnarray}
In the model Hamiltonian the kinetic energy is the sum of intersite hopping and onsite interorbital hopping, and the Coulomb interactions will exist between any two electrons so that it should be a sum of intersite and onsite Coulomb interactions. However, in usual studies only the onsite Coulomb interactions are concentrated, and the intersite Coulomb interaction are treated as potential with mean field approximation and dropped in the Hamiltonian.
If directly solving the cluster model or studying higher order spacial fluctuations, the intersite Coulomb interactions should also be included into the Hamiltonian.

While in the Kanamori's form,\cite{Kanamori63} the multiorbital Hamiltonian is written as
\begin{eqnarray}
{\cal H}&=&-\sum_{ijlm\sigma,i\neq j}t_{ijlm}f^{\dag}_{il\sigma}f_{jm\sigma}
+\sum_{il}U_{ll}\hat{n}_{il\uparrow}\hat{n}_{il\downarrow}\nonumber\\
&&+\sum_{ilm\sigma\sigma',l<m}U_{lm\sigma\sigma'}\hat{n}_{il\sigma}\hat{n}_{im\sigma'}\nonumber\\
&&-J\sum_{iml,m\neq l}f^{\dag}_{il\sigma}f_{il\sigma'}f^{\dag}_{im\sigma'}f_{im\sigma}\nonumber\\
&&+J\sum_{iml,m\neq l}f^{\dag}_{il\sigma}f^{\dag}_{il\sigma'}f_{im\sigma'}f_{im\sigma},
\end{eqnarray}
where the last two terms are the so-called spin-flip term and pair-hopping term. Or even some people
write the Hamiltonian in a more compact form
\begin{eqnarray}
{\cal H}&=&-\sum_{ijlm\sigma,i\neq j}t_{ijlm}f^{\dag}_{il\sigma}f_{jm\sigma}\nonumber\\
&&+\sum_{ilm1234}U^{lm}_{1234}f^{\dag}_{il1}f_{il2}f^{\dag}_{im3}f_{im4}.
\end{eqnarray}
The spin-flip term and pair-hopping term are usually assumed to be some kinds of Coulomb interactions and $J$ has also the same unit of Coulomb interaction strength $U$. However, the Coulomb interactions can only occur between two electrons. In the above equations $\hat{n}_{fm\sigma}=f^{\dag}_{m\sigma}f_{m\sigma}$ is the occupation operator of electrons with $\sigma$ spin in $m$th orbital so that $U\hat{n}_{fm\sigma}\hat{n}_{fl\sigma'}$ means Coulomb interaction between two electrons, where $l$ can equal $m$ and $\sigma$ can equal $\sigma'$ but can not equal simultaneously. If the spin-flip and pair-hopping terms are considered to be Coulomb interactions, $f^{\dag}_{l\sigma}f_{m\sigma}$ must have the physical meaning of occupation. However, it labels the onsite interorbital single-electron hoppings. Although the matrix element of $f^{\dag}_{l\sigma}f_{m\sigma}$ between two basis functions may be nonzero, one should not consider they are Coulomb interaction terms. One can also notice that from our introduction of Eq.~\eqref{eq6} in energy of the system there is also no position for these two terms so that they should not be added into Hamiltonian. A reasonable way to introduce the spin-flip term and pair-hopping term is from the onsite interorbital single-electron hoppings and consider them as some kinds of quantum effect. Apart from the onsite single electron hopping. The spin-flip and pair hopping terms do relate to the Coulomb interactions. The physical meaning of these two terms is that they are some kinds of higher order correlation effects between two occasional single-electron hoppings, where these two hoppings are coupled or correlated with Coulomb interactions.

Some people may argue that by choosing special basis the onsite interorbital single-electron hoppings may be eliminated, which has been discussed in Ref.~\onlinecite{FO11c} but may still exist some confusion. Actually this is not always true because usually the orthonormal basis are chosen to make the matrix element of $<\phi_{\mu}|f^{\dag}_{l\sigma}f_{m\sigma}|\phi_{\nu}>$ to be zero, while $<\phi_{\mu}|f^{\dag}_{l\sigma}f_{m\sigma}|\phi_{\nu}>$ is only one of the consequence of the single-electron hopping term but not the interaction itself. Usually the chosen orthonormal basis can not guarantee the expectation value of both single-electron interorbital hopping term and its two-particle offsprings simultaneously to be zero. That's why the spin-flip term and pair-hopping can survive in orthonormal basis.
The relation between the onsite interorbital single-electron hoppings and the spin-flip term is that in EOMs or diagram expansion the spin-flip term can be reproduced by the single-electron hoppings. So does the pair-hopping term and other two-particle correlations shown in Eq.~(9) in Ref.~\onlinecite{FO11c}.
If all the onsite interorbital single-electron hoppings vanish (i.e., the interaction strengths are all zero due to the physical situation of the system, e.g., strictly prohibited by the quantum rules, or completely destroyed by the choosing of the basis which is a hard or even impossible task when there are many partially occupied orbitals), the spin-flip term and pair hopping term will also vanish and the Hamiltonian will return to
\begin{eqnarray}
{\cal H}&=&-\sum_{ijlm\sigma,i\neq j}t_{ijlm}f^{\dag}_{il\sigma}f_{jm\sigma}
+\sum_{il}U_{ll}\hat{n}_{il\uparrow}\hat{n}_{il\downarrow}\nonumber\\
&&+\sum_{ilm\sigma\sigma',l<m}U_{lm\sigma\sigma'}\hat{n}_{il\sigma}\hat{n}_{im\sigma'}
\end{eqnarray}
instead of Hamiltonian in Kanamori's form. On the contrary once spin-flip term and pair-hopping term exist, there must be nonzero onsite interorbital single-electron hoppings existing, which can be achieved from the EOMs in Ref.~\onlinecite{FO11c}. When the number of orbital is larger than two, these additional two-particle correlation terms in Eq.~(9) in Ref.~\onlinecite{FO11c} have identical amplitudes (or possibility) with the spin-flip term and pair hopping term. In a self-consistent manner, if one wants to study the effects of spin-flip term and pair-hopping term, these additional terms in same order can not be neglected. Moreover, one may notice that the Hamiltonian in Eq.~\eqref{eq6} can maximally have $SU(N)$ symmetry in special physical conditions where $N$ is the total number of particles in the system. As a more general multiorbital model, it can of course to describe a special system with only $SU(2)$ symmetry by manipulating the prefactors of those interorbital single electron hopping terms.

In summary the EOM method is a powerful tool to explore a correlated many body system. It can take into account the many body correlation effects in deriving the higher and higher order EOMs, but not based on the perturbation nor adding higher order correlations into Hamiltonian. From this point, it can help to build many body theory self-consistently and can also have more advanced implementations in the studies of correlated many body systems.

The author would like to thank for the financial support from the Swedish Foundation for Strategic Research SRL grant 10-0026.

\end{document}